\documentclass[conference,compsoc]{IEEEtran}
\ifCLASSOPTIONcompsoc
  \usepackage[nocompress]{cite}
\else
  \usepackage{cite}
\fi
\usepackage{lmodern}
\usepackage{etoolbox}
\usepackage{graphicx}
\usepackage{titlesec}
\usepackage{tcolorbox}
\usepackage{array}
\usepackage{booktabs}
\usepackage{amsmath}
\usepackage{enumitem}
\usepackage{eucal}
\usepackage[symbol]{footmisc}

\title{\Huge \textbf{5G Network Automation Using Local Large Language Models and Retrieval-Augmented Generation}}

\author{
\IEEEauthorblockN{\fontseries{b}\selectfont Ahmadreza Majlesara\IEEEauthorrefmark{1}, Ali Majlesi\IEEEauthorrefmark{1}, Ali Mamaghani\IEEEauthorrefmark{1}, Alireza Shokrani\IEEEauthorrefmark{1}, Babak Hossein Khalaj\footnote{Corresponding author.}\IEEEauthorrefmark{1}}
\IEEEauthorblockA{\textit{\IEEEauthorrefmark{1}Department of Electrical Engineering, Sharif University of Technology, Tehran, Iran} \\
Email: \{ahmad.ara, ali.majlesi, ali.mamaghani, alireza.shokrani, khalaj\}@sharif.edu}
}

\titleformat{\section}[block]{\centering\bfseries\MakeUppercase}{\thesection.}{1em}{}
\titleformat{\subsection}[block]{\centering\bfseries}{\thesubsection.}{1em}{}
\renewcommand{\thesection}{\Roman{section}}
\renewcommand{\thesubsection}{\Roman{section}.\Alph{subsection}}

\begin{document}

\maketitle

\begin{abstract}
  This demonstration showcases the integration of a lightweight, locally deployed Large Language Model (LLaMA-3 8b Q-4b) empowered by retrieval-augmented generation (RAG) to automate 5G network management, with a strong emphasis on privacy. By running the LLM on local or edge devices, we eliminate the need for external APIs, ensuring that sensitive data remains secure and is not transmitted over the internet. Although lightweight models may not match the performance of more complex models like GPT-4, we enhance their efficiency and accuracy through RAG. RAG retrieves relevant information from a comprehensive database, enabling the LLM to generate more precise and effective network configurations based on natural language user input. This approach not only improves the accuracy of the generated configurations but also simplifies the process of creating and configuring private networks, making it accessible to users without extensive networking or programming experience. The objective of this demonstration is to highlight the potential of combining local LLMs and RAG to deliver secure, efficient, and adaptable 5G network solutions, paving the way for a future where 5G networks are both privacy-conscious and versatile across diverse user profiles.
\end{abstract}

\IEEEpeerreviewmaketitle

\section{Introduction}
The concept of private networks has come into importance with the global shift in mobile technology from 4G to 5G and onwards. Private  networks provide bespoke connectivity options, but setting these kind of networks up often requires broad knowledge in networking and software development. At the same time, artificial intelligence developments, particularly large language models, are also revolutionizing communication infrastructures by offering new ways to automate complex technical tasks easily. Thus, the idea of using LLMs to realize automation in configuration and management tasks in private networks has been discussed and demonstrated in recent years. The issue has been addressed and elaborated on in the work titled LLM for 5G: Network Management.\cite{EURECOM+7687}\\

This automation, although, comes with some overhead in implementation. the main challenges investigated in this demo are privacy and cost efficiency. Generally, LLMs deployed on the cloud, such as ChatGPT, need API access and the transmission of data across the internet, raising privacy issues. Besides, all these cloud solutions are pretty pricey because they charge for API usage. This was, however, brought to light in recent times with the advent of lightweight and efficient LLMs that could run on the local machine.The models, apart from their core functionality, have the dual advantage of reduced operational costs -that can be done by accessible GPUs- and heightened privacy by not having to send sensitive data over the internet to external servers.Although local models do not undergo pre-training for advanced tasks and demonstrate lower accuracy compared to more sophisticated models such as GPT-3.5, the capacity to implement these models within local settings undoubtedly represents a significant advancement in improving safety and effectiveness in the management of networks.\\

The introduction of Retrieval-Augmented Generation (RAG) has considerably improved AI-driven communication systems by allowing large language models (LLMs) to retrieve pertinent information as required, thereby generating outputs with enhanced precision. The capacity of RAG to utilize current databases minimizes the necessity for ongoing fine-tuning and results in reduced expenses. Our new system introduces several key innovations:\\

\begin{itemize}[leftmargin=*]
\item \textbf{Novel Use of Local LLMs:} We leverage locally deployed LLMs, ensuring that all data processing occurs within secure, isolated environments, thereby enhancing privacy and security. Furthermore, this approach reduces the cost of operation.

\item  \textbf{Integration of RAG:} By incorporating a general-purpose RAG model, our system allows for real-time retrieval of relevant data, enabling on-the-fly adaptation to new requirements and scenarios. The adaptive database architecture facilitates seamless updates, making the system both flexible and efficient.\\

\end{itemize}

\begin{figure}[t]
  \centering
  \includegraphics[width=0.5\textwidth]{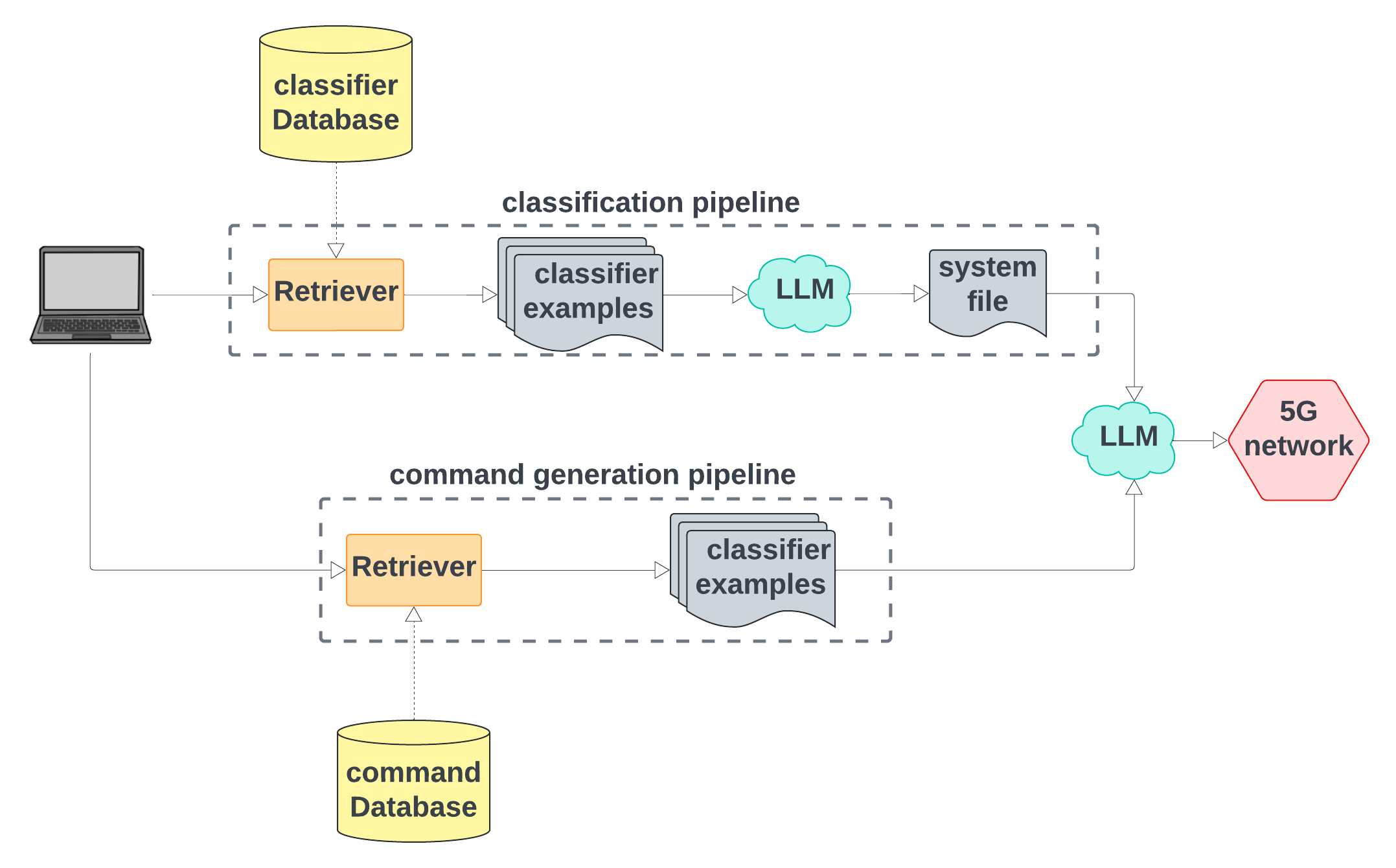}
  \caption{this illustration shows our system overview and flow of data. user instruction in natural language is given to 2 parallel pipelines to get system files and retrieve similar samples from databases and give these to the LLM to generate the commands that user requested.}
  \label{fig:system_overview}
  
\end{figure}

These contributions represent a significant advancement in the automation of private 5G network creation, making the process more accessible, cost-efficient, and secure. By integrating state-of-the-art AI technologies in a privacy-conscious manner, our work paves the way for broader adoption of private networks in future communication infrastructures.

\section{Methodology}
The LLM model generates commands from a user's natural language description in two steps:
\begin{enumerate}
    \item Command Classification
    \item Command Generation
\end{enumerate}
In the first step, LLM detects the class of the command based on the user's prompt. In the second step, a description of the specified class and samples retrieved from a corpus are provided to LLM to generate the command.

These two steps will be discussed in the following subsections.
\subsection{Step 1: Command Classification}
In the first step, the LLM identifies the class of the command, which can be one of 11 possible classes, such as "test," "create," or "remove." To assist with this classification, first a sample retriever  finds the 8 most similar examples from the dataset. These examples, along with their class labels, are appended to the end of the prompt (see Fig. \ref{fig:calssifier_prompt}). The LLM then uses this prompt to determine the appropriate command class. The retrieved examples ensure that the LLM generates the command in the correct syntax\cite{gao2020making}. For the retrieval process, we used BAAI/bge-small-en-v1.5\cite{bge_embedding} as the embedding model and the Llama Index retriever\cite{Liu_LlamaIndex_2022}.

\subsection{Step 2: Command Generation}
Once the command class is determined, the LLM generates the specific command in the second step. The system prompt associated with the selected command class is used to form the prompt for this step. This system prompt includes a description of the command’s task, its options, and any relevant flags. The retriever then retrieves similar examples from the command corpus and appends them to the prompt. The LLM uses this information to generate the command.\\

After the LLM generates the response which include the description and command, it is filtered using regular expressions to extract only the command itself, removing any extra explanations or details.

\section{Model Implementation}
We used a 4-bit quantized version of Llama 3 with 8 billion parameters as the LLM. This model was implemented using Ollama\cite{ollama}, a Python library that allows LLMs to be run locally. Remarkably, this implementation requires less than 6GB of vRAM, making it feasible to run on laptops equipped with NVIDIA 3060 GPUs, which are common today.

\begin{figure}
    \centering

{\begin{tcolorbox}[colframe=black,
                  boxrule=0.5pt,
                  arc=2mm,
                  top=2mm, bottom=2mm, left=1mm, right=1mm, 
                  boxsep=5pt]
\raggedright
System Prompt:\\
You are a classifier to classify the input command into the categories below:\\
\begin{enumerate}
    \item user: add a new user from its IMSI number to the network.
    \item list: List all of the users, gnode-bs, or nodes.\\
    ...
\end{enumerate}
------------------------------------\\
Instruct:

Could you please give me the list of active users

Samples:

\begin{enumerate}
    \item Input: Could you kindly offer me a the list of active users since 2024/08/10
    
    Output: list
    
    \item Input: I want list of active users

    Output: list

    ...
\end{enumerate}
------------------------------------\\
Answer:\\
list
\end{tcolorbox}
}
    \caption{An example of the classifier prompt}
    \label{fig:calssifier_prompt}
\end{figure}

\begin{figure}
    \centering

{\begin{tcolorbox}[colframe=black,
                  boxrule=0.5pt,
                  arc=2mm,
                  top=2mm, bottom=2mm, left=1mm, right=1mm, 
                  boxsep=5pt]
\raggedright
System Prompt:

You are assistant to … The input is a message in which someone is trying to get list of users, nodes, gnodebs. …

Base command 1: f the user wants the list of users, this is the base command:

“list users ”

If the list of active users between  \texttt{<start\_time>} and \texttt{<end\_time>} requested also add - active \texttt{<start\_time> <end\_time>} flag, similar to this command:

“list users –active 20240801 20240901”

Which gives active users from 2024/8/1 to 2024/9/1,

If the user does not specify the time interval or requested already active users use –active now flag, similar to:

“list users –active now”

If the user does not specify the end time but says the start time use -active \texttt{<start\_time> now} flag similar to:

“list users –active 20240801 now”\\
------------------------------------\\
Instruct:\\
Could you please give me the list of active users since 2 March.\\
Samples:

\begin{enumerate}
    \item \texttt{Input} : Could you kindly offer me a the list of active users since 2024/08/10 ?\\ \texttt{Output :\\ list users} –\texttt{active 20240810 now}\\
    ...\\
    \item \texttt{Input} : ... ,\texttt{Output} : ... \\
    ...\\
\end{enumerate}
------------------------------------\\
Answer:\\
\texttt{list users }–\texttt{active 20240301 now}
\end{tcolorbox}
}
    \caption{An example of the command generator's prompt}
    \label{fig:genrator_prompt}
\end{figure}

\section{Results}
The model's performance was evaluated using two metrics:
\begin{enumerate}
    \item Accuracy
    \item Uni-gram precision
\end{enumerate}

\textbf{Accuracy} was determined by comparing the model's entire output with the corresponding ground truth in the dataset. A sample was considered correct if its output was identical to the ground truth; otherwise, if there was even a single word difference, the sample was marked as incorrect.

\textbf{Uni-gram precision} measures the proportion of correct tokens in the output relative to the total number of output tokens. This metric is calculated as:

% TODO : Breaking the eqn
\begin{equation}
\begin{split}
&\text{uni-gram\;precision} =\\
&\frac{| \text{output tokens} \cap \text{ground truth tokens}|}{|\text{output tokens}|}
\end{split}
\end{equation}

Here, $|\mathcal{A}|$ represents the number of elements in the set $\mathcal{A}$. Uni-gram precision was chosen as a metric because, in some cases, the order of flags in commands generated by the LLM differs from the order in the ground truth commands. Since the order of these flags is not important, this metric provides a more meaningful evaluation of the output quality.

The results indicate that incorporating a sample retriever improves the uni-gram precision of the local Llama 3 model to 68\%, representing a notable 18\% increase in this metric (as shown in Table \ref{tab:RAGS}). Also this method increases accuracy up to 46\% which is 25\% more than non-RAG model.
\begin{table}[t]
\caption{Accuracy with different LLM}
\centering
\small 
\renewcommand{\arraystretch}{1.2} 
\setlength{\tabcolsep}{3pt} 
\begin{tabular}{>{\raggedright\arraybackslash}p{3cm} 
                  >{\raggedright\arraybackslash}p{1cm}
                  >{\raggedright\arraybackslash}p{1.5cm}
                  >{\raggedright\arraybackslash}p{2cm} }
\toprule
\textbf{LLM Model} & \textbf{RAG} & \textbf{Accuracy}& \textbf{uni-gram precision} \\
\midrule
Fine tuned GPT 3.5 & no & 69.5\% & 95\% \\
GPT 4 & no & 44.3\% & 52.5\%\\
Llama 3 8b & no & 21\% & 47.4\%\\
Llama 3 8b & yes & 46.1\% & 68.1\%\\
\bottomrule
\end{tabular}
\label{tab:RAGS}
\vspace{-0.5em}
\end{table}

\section{Conclusion}
This work proposes a framework that integrates a local LLM with a RAG system to enable the automatic configuration of a private 5G network through natural language requests. The framework achieves a uni-gram precision of 68\%, reflecting an 18\% improvement over the non-RAG system.
\section{Future Works}
% Add the referece of LoRA
While RAG techniques improve uni-gram precision by up to 18\%, this is still significantly lower than the 95\% uni-gram precision achieved by the fine-tuned GPT-3.5 model. To narrow this gap, the Llama 3 model could be fine-tuned using Parameter-Efficient Fine-Tuning methods, such as LoRA\cite{hu2021lora}.

\bibliographystyle{IEEEtran}
\bibliography{citation}

\end{document}